\documentclass[11pt]{article}

\usepackage{titling}
\usepackage{amsmath,amssymb,amsfonts,amsthm}
\usepackage{authblk}
\usepackage{blkarray}
\usepackage{booktabs}
\usepackage{caption}
\usepackage{enumitem}
\usepackage[margin=1in]{geometry}
\usepackage{graphics,graphicx,epsfig,float,geometry}
\usepackage{ltablex}
\usepackage[round]{natbib}
\usepackage{rotating}
\usepackage{setspace}
\usepackage[]{subcaption}
\usepackage{tabularx}
\usepackage{tgpagella}
\usepackage{titlesec}
\usepackage{xfrac}
\usepackage{color}

\titleformat*{\section}{\normalsize\bfseries}
\titleformat*{\subsection}{\normalsize\bfseries}
\titleformat*{\subsubsection}{\normalsize\bfseries}
\titleformat*{\paragraph}{\normalsize\bfseries}
\titleformat*{\subparagraph}{\normalsize\bfseries}

\numberwithin{equation}{section}
\DeclareMathOperator*{\argmax}{arg\,max}

\title{The gig economy during an epidemic: coupling disease transmission with labour market dynamics}
\author[1]{Bryce Morsky}
\author[2]{Tyler Meadows}
\author[2]{Felicia Magpantay}
\author[2]{Troy Day}
\affil[1]{Department of Mathematics, Florida State University, Tallahassee, FL, USA}
\affil[2]{Department of Mathematics \& Statistics, Queen's University, Kingston, ON, CA}
\date{\today}

\begin{document}

\maketitle

\begin{abstract}
    The gig economy has grown significantly in recent years, driven by the emergence of various facilitating platforms. Triggering substantial shifts to labour markets across the world, the COVID-19 pandemic has accelerated this growth. To understand the crucial role of such an epidemic on the dynamics of labour markets of both formal and gig economies, we develop and investigate a model that couples disease transmission and a search and match framework of unemployment. We find that epidemics increase gig economy employment at the expense of formal economy employment, and can increase the total long term unemployment. In the short run, large sharp fluctuations in labour market tightness and unemployment can occur, while in the long run, employment is reduced under an endemic disease equilibrium. We analyze a public policies that increase unemployment benefits or provide benefits to gig workers to mitigate these effects, and evaluate their trade-offs in mitigating disease burden and labour market disruptions.
\end{abstract}

\noindent{\textbf{Keywords:} dual labour markets, epidemics, gig economy, search and match, unemployment}\\
\noindent{\textbf{JEL:} I18, J46, J64}

\section{Introduction}

The COVID-19 pandemic significantly disrupted labour markets worldwide, presenting extraordinary economic challenges to firms, workers, and policy makers \citep{costadias20}. It led to widespread layoffs and furloughs, creating a swell of temporary unemployment \citep{gallant20}. These shocks have raised concerns about long term unemployment and structural changes to labour markets \citep{sharone21}. For example, remote work became more common and normative \citep{cook20,newbold22,vyas22}, and there was an increased participation of both workers and customers in the gig economy \citep{katz19,apouey20,joo21,umar21}, where work is non-standard, casual, and independent \citep{woodcock20}.

The gig economy has been significantly growing even before the COVID-19 pandemic. For example, by 2016, gig workers had risen to $8.2$\% of the share of all workers in Canada \citep{jeon21}. The gig economy has particularly expanded in regions where unemployment was high \citep{rozzi18}. ``Side hustles" provided a social safety net \citep{ravenelle21} and temporary and low income employment during the pandemic \citep{hasegawa22}. Labour tightness increased in the short term as workers reduced their search for jobs or found jobs less negatively affected by the epidemic \citep{hensvik21}. Workers and firms also changed their job search patterns \citep{bauer21,bauer23}. This resulted in an expansion of searches in new areas \citep{carrillo23}, as well as a rise in mismatch between employees and jobs \citep{pizzinelli23}. However, these effects on firms and workers varied greatly: the pandemic harmed some and benefited others \citep{batool21}.

Previous research has investigated the interplay of epidemics and labour markets, particularly economic behavioural responses to an unfolding epidemic \citep{boppart20,barbierigoes21,bradley21,eichenbaum21}. These biological-economic interactions have been explored both indirectly through shocks or productivity loss \citep{gregory20,guerrieri22} and directly by explicitly coupling models of disease progression and employment \citep{kapicka22,glover23,jackson24}. Such models have shown that labour markets can be inefficient during epidemics \citep{kapicka22} and can result in worker skill loss and thereby long-term productivity loss \citep{jackson24}. Although public policy can mitigate this effect during the pandemic \citep{kapicka22,guerrieri22,glover23}, quarantines, in particular, can lead to long-term productivity loss \citep{gregory20}. These models frequently feature rational actors trying to balance the trade-offs in risk of infection and monetary benefits. Such individual decision-making and utility maximization also play a critical role in epidemic models. Individuals weigh the costs and benefits of becoming ill, vaccines, non-pharmaceutical interventions, and other social, biological, and economic factors \citep{bauch05,bauch12,brotherhood20,garibaldi20,eichenbaum21,qiu22,morsky23,morsky25}.

To understand how an epidemic can shape unemployment and employment in formal and gig labour markets, both in the short term and the long term, we extend the canonical search and matching model of unemployment \citep{pissarides85,mortensen94,pissarides00} to incorporate the spread of disease into the dynamics of two labour markets. The first labour market represents the ``non-gig" or ``formal" labour market while the second represents the labour market for the gig economy. We assume that employees in the gig economy are less productive than the formal economy, but have a lower chance of being infected. With respect to the epidemiological dynamics, we consider waning immunity and thus the epidemic can become endemic with sustained effects on the labour markets. We also explore how exogenous factors such as public policies to reduce the spread of infections impact both labour markets.

\section{Methods}

Here we develop the model that couples economic and epidemiological dynamics. Parameter and variable values, their definitions, and calibrated values are presented in Table \ref{tbl:param}. We begin with an exposition of the population dynamics followed by that of the value functions for firms and employees. Section \ref{sec:calibration} details the calibration of the default values.

\begin{table}[ht!]
\begin{center}
\begin{tabular}{lll}
\toprule
Parameter & Definition & Default values \\
\midrule
$\alpha$ & workers' bargaining power &  $0.5$ \\
$\beta_f$ & transmission rate for employed workers & $0.4$/day \\
$\beta_g$ & transmission rate for gig workers & $0.38$/day \\
$\beta_u $ & transmission rate for unemployed & $0.36$/day \\
$1/\gamma$ & recovery period & $5$ days \\
$\eta$ & elasticity of the matching function & $0.5$\\
$\theta_f$ & pre-pandemic tightness in the formal labour market & $1$\\
$\theta_g$ & pre-pandemic tightness in the gig labour market & $5.692$\\
$\lambda_f$ & job loss rate & $ 0.001096$/day\\
$\lambda_g $ & job loss rate & $0.0054794$/day \\
$\mu$ & matching function constant & $0.01517$ \\
$1/\rho$ & resistance period & $100$ days \\
$b_u $ & unemployment benefit & $0.4$/day\\
$b_g $ & gig work flexibility benefit & $0.2$/day \\
$c_f$ & vacancy cost in the formal economy & $0.1731$\\
$c_g$ & vacancy cost in the gig economy & $0.07253$ \\
$r$ & interest rate & $ 0.0001337$/day \\
$W_f$ & pre-pandemic formal wage  & $1$/day\\
$W_g $ & pre-pandemic gig wage &$ 0.85$/day\\
$y_f$ & worker productivity in the formal economy & $1.01407$/day \\
$y_g$ & worker productivity in the gig economy&$0.9141$/day \\
\bottomrule
\end{tabular}
\caption{Parameters and the default values used for numerical simulations.} \label{tbl:param}
\end{center}
\end{table}

\subsection{A compartmental model of disease spread}

The population is split into proportions that are susceptible, infectious, and recovered, $S$, $I$, and $R$, respectively. These categories are further divided into those who are employed in the formal and informal (gig) labour markets, and those who are unemployed, with subscripts $f$, $g$, and $u$, respectively. Therefore,
\begin{align}
    S &= S_f+S_g+S_u, \\
    I &= I_f+I_g+I_u, \\
    R &= R_f+R_g+R_u,
\end{align}
and $S+I+R=1$. We develop a system of differential equations to model the flow --- both from epidemiological and economic factors --- among these compartments. First consider the epidemiological aspects of the model. We assume that infectious individuals behave similarly regardless of employment status. Thus, the transmission rates of the disease, $\beta_j$ for $j \in \{ f, g, u\}$, depend solely on the employment status of susceptible individuals with $\beta_f > \beta_g > \beta_u$. We make this assumption since work-related transmission of COVID-19 was significant \citep{lan20}, while gig work, in contrast, is casual and contingent in nature. The constant $\gamma$ is the recovery rate and $\rho$ is the per capita rate at which immunity wanes.

The economic aspects of the model are developed from a search and matching model of unemployment. The constants $\lambda_f$ and $\lambda_g$ are the rates at which a worker employed in formal or gig work, respectively, loses their job due structural shifts or shocks. We assume that $\lambda_j$ is independent of the infection status of the employee and the state of the epidemic. Frictions in the labour market create unemployment and are determined by a matching function that pairs unemployed individuals with vacancies. This matching function models employee recruitment, work search, evaluation of candidates, etc.--- all things that can cause frictions in the labour market. The ratio of this matching function to the vacancy rate, i.e.\ the per capita rate of filling a vacancy, in labour market $j$ is $q(\Theta_j)$ where $\Theta_j$ is the labour market tightness, i.e.\ the ratio of vacancies divided by unemployment in the labour market for economy $j$. The ratio of the matching function to the proportion of unemployed individuals is $p(\Theta_j) = \Theta_jq(\Theta_j)$, which is the rate at which an unemployed individual finds a job. Here will will consider such functions derived from a Cobb-Douglas matching function:
\begin{align}
    q(\Theta_j) = \mu\Theta_j^{-\eta}, \\
    p(\Theta_j) = \Theta_j q(\Theta_j) = \mu\Theta_j^{1-\eta}
\end{align}
where $\mu>0$ and $\eta \in (0,1)$.

We assume that employers do not know whether or not potential employees are susceptible or recovered: indeed, it is unlikely that individuals themselves would have this information due to uncertainty in the duration of one's waning immunity. Thus, the matching function is the same for both susceptible and recovered individuals. However, infectious individuals are unable to search for work (this was also assumed in \cite{kapicka22} and does is not fundamentally affect our results). The equations for the dynamics of workers in the formal and gig labour markets ($j \in \{ f,g \}$) are thus:
\begin{align}
    \dot{S}_j & =  \rho R_j - \beta_j S_j I + p(\Theta_j)S_u - \lambda_j S_j, \label{Sj_dot} \\
    \dot{I}_j & =  \beta_j S_j I - \gamma I_j - \lambda_j I_j, \label{Ij_dot} \\
    \dot{R}_j & =  \gamma I_j - \rho R_j + p(\Theta_j)R_u - \lambda_j R_j \label{Rj_dot}.
\end{align}
And the dynamics for the unemployed are:
\begin{align}
    \dot{S}_u & =  \rho R_u - \beta_u S_u I - (p(\Theta_f) + p(\Theta_g))S_u + \lambda_f S_f + \lambda_g S_g, \label{Su_dot} \\
    \dot{I}_u & =  \beta_u S_u I - \gamma I_u + \lambda_f I_f + \lambda_g I_g, \label{Iu_dot} \\
    \dot{R}_u & =  \gamma I_u - \rho R_u - (p(\Theta_f) + p(\Theta_g))R_u + \lambda_f  R_f + \lambda_g R_g. \label{Ru_dot} 
\end{align}
Let $F=S_f+I_f+R_f$ be the proportion of individuals employed in the formal economy. Similarly defined, let $G$ and $U$ be the proportions of gig workers and unemployed. Summing the disease status dynamics for each, the dynamics for the employed and unemployed are:
\begin{align}
    \dot{F} &= p(\Theta_f)(S_u+R_u) - \lambda_f F, \\
    \dot{G} &= p(\Theta_g)(S_u+R_u) - \lambda_g G, \\
    \dot{U} &= \lambda_f F + \lambda_g G - (p(\Theta_f)+p(\Theta_g))(S_u+R_u).
\end{align}
Note that we assume that employees must become unemployed before they can become employed elsewhere. i.e.\ there is no on-the-job searching for employment.

The outcome of this model depends on the labour market tightness for each labour market, which in turn will depend upon the value functions for employees and employers. These value functions are affected by the spreading of disease and the average infection status of workers. In the next section, we develop these value functions.

\subsection{The value functions, wage determination, and labour market tightness}

An employee in the formal economy produces an exogenous constant amount of production worth $y_f$ per unit of time while an employee in the gig economy produces an amount $y_g$ per unit time. We assume that $y_f > y_g$. Each employee receives a wage $W_f$ or $W_g$, which is time varying and endogenous, in the formal and gig economies, respectively. We also assume that there is an exogenous constant income for unemployed workers $b_u$ provided by the government,and that gig workers receive an additional intangible benefit $b_g$ due to the nature of their work (additional time off, more flexible hours, etc.).

The present discounted value of being employed in economy $j$ is $V_j$, and being unemployed is $V_u$. $\dot{V}_j$ is a potential capital gain of being in state $j$ while the system is not in equilibrium. Given a discount rate $r$, the resulting Bellman equations are:
\begin{align}
    rV_f &= W_f + \dot{V}_f - \lambda_f(V_f - V_u), \label{value_employed_formal} \\
    rV_g &= W_g(1-i_g) + b_g + \dot{V}_g - \lambda_g(V_g - V_u), \label{value_employed_gig} \\
    rV_u &= b_u + \dot{V}_u - p(\Theta_f)(V_u - V_f) - p(\Theta_g)(V_u - V_g). \label{value_unemployed}
\end{align}
We assume that sick employees in the formal labour market go on sick leave, receiving their full wage, and that the other negative effects of the disease are negligible. Therefore, they receive a dividend $W_f$. However, gig workers do not go on sick leave and thus do not earn wages when ill. Letting $i_j=I_j/(S_j+I_j+R_j)$ be the proportion of infectious workers in $j$, then $W_g(1-i_g)+b_g$ is the expected dividend for gig workers. Employees thus factor in the chance of being infected into the dividend, weighting the costs of being infected. Since unemployed individuals receive no wage and the symptoms of the disease are mild, they receive a dividend of $b_u$ regardless of their infection status. Note that we assume that individuals do not consider their current infection status when determining the value of being employed in economy $j$, since reinfections are common and individuals may be uncertain about their susceptibility.

Each job, whether from the formal or gig economies, can be either filled with a worker or vacant. There is a constant exogenous cost $c_j$ to maintain a job in economy $j$, which exists whether or not the job is filled. Gig jobs tend to have lower fixed hiring costs, since they rely on platforms such as Uber that are scalable and automate much of the vacancy posting and matching process. Further, gig jobs tend to have lower onboarding costs and fewer legal obligations than formal jobs. Therefore, we assume that $c_f > c_g > 0$. The profits an employee produces for a firm in the formal economy depend on the employee's infection status. With frequency $1-i_j$, the employee is uninfected and produces a dividend of $y_j-W_j$ for the employer. Infectious employees, however, produce a dividend $-W_f$ in the formal economy (they are paid a wage yet are unproductive) while they produce no dividend in the gig economy (they are unproductive yet unpaid). New jobs can be created at any time and thus the total number of jobs in both economies is endogenous. The present discounted values to a firm for a filled and unfilled vacancy are $V_{f,j}$ and $V_{v,j}$, respectively. These must satisfy the Bellman equations:
\begin{align}
    rV_{f,f} &= y_f(1-i_f) - W_f + \dot{V}_{f,f} - \lambda_f(V_{f,f} - V_{v,f}), \label{value_filled_formal} \\
    rV_{f,g} &= (y_g - W_g)(1-i_g) + \dot{V}_{f,g} - \lambda_g(V_{f,g} - V_{v,g}), \label{value_filled_gig} \\
    rV_{v,j} &= - c_j + \dot{V}_{v,j} - q(\Theta_j)(V_{v,j} - V_{f,j}). \label{value_unfilled}
\end{align}
Free entry implies that $V_{v,j} = \dot{V}_{v,j} = 0$, which in turn implies that:
\begin{align}
    \dot{V}_{f,f} &= W_f -y_f(1-i_f)+(r+\lambda_f)V_{f,f}, \label{filled_vacancy_diff_formal} \\
    \dot{V}_{f,g} &= (W_g-y_g)(1-i_g) + (r+\lambda_g)V_{f,g}, \label{filled_vacancy_diff_gig} \\
    V_{f,j} &= \frac{c_j}{q(\Theta_j)}. \label{filled_vacancy_eq}
\end{align}
Further, assuming that firms can enter either labour market, then $V_{f,f} = V_{f,g}$.

To determine the equilibrium wage, we need to optimize
\begin{equation}
    W_j = \argmax_{W_j} (V_j-V_u)^\alpha(V_{f,j} - V_{v,j})^{1-\alpha},
\end{equation}
where $\alpha$ is the worker's bargaining power. The first order condition for optimality gives us
\begin{equation}
    V_j - V_u = \frac{\alpha}{1-\alpha}(V_{f,j} - V_{v,j}). \label{wage_opt_cond}
\end{equation}
Subtracting Equation \ref{value_unemployed} from Equations \ref{value_employed_formal} and \ref{value_employed_gig} and rearranging we have
\begin{align}
    (r+\lambda_f)(V_f-V_u) - (\dot{V}_f-\dot{V_u}) &= W_f - b_u + p(\Theta_f)(V_u-V_f) + p(\Theta_g)(V_u-V_g), \label{wage_eqn1_formal} \\
    (r+\lambda_g)(V_g-V_u) - (\dot{V}_g-\dot{V_u}) &= W_g(1-i_g) + b_g - b_u + p(\Theta_f)(V_u-V_f) + p(\Theta_g)(V_u-V_g). \label{wage_eqn1_gig}
\end{align}
By substituting Equation \ref{wage_opt_cond} and its derivative with respect to time along with Equations \ref{filled_vacancy_eq}, \ref{filled_vacancy_diff_formal}, and \ref{filled_vacancy_diff_gig} into Equations \ref{wage_eqn1_formal} and \ref{wage_eqn1_gig} and rearranging we have the wage equations:
\begin{align}
    W_f &= \alpha y_f(1-i_f) + (1-\alpha)b_u + \alpha c_f\Theta_f + \alpha c_g\Theta_g,
    \label{wage_eqn_formal} \\
    W_g(1-i_g) &= \alpha y_g(1-i_g) + (1-\alpha)(b_u - b_g) + \alpha c_f\Theta_f + \alpha c_g\Theta_g. \label{wage_eqn_gig}
\end{align}

We may now calculate equilibrium labour market tightness. Substituting Equations \ref{filled_vacancy_eq}, \ref{wage_eqn_formal}, and \ref{wage_eqn_gig} into Equations \ref{filled_vacancy_diff_formal} and \ref{filled_vacancy_diff_gig}, differentiating with respect to time, and setting $\dot{\Theta}_j=0$ gives us
\begin{align}
    &(1-\alpha)(y_f(1-i_f) - b_u) -\alpha c_f\Theta_f - \alpha c_g\Theta_g - \frac{c_f(r+\lambda_f)}{q(\Theta_f)} = 0, \label{theta_formal} \\
    &(1-\alpha)(y_g(1-i_g) + b_g - b_u) -\alpha c_f\Theta_f - \alpha c_g\Theta_g - \frac{c_g(r+\lambda_g)}{q(\Theta_g)} = 0, \label{theta_gig}
\end{align}
which completes the model.

\section{Results}

\subsection{Steady state analysis} \label{sec:steady_state}

Here we analyze the steady states of the system, namely when $\dot{S}_j=\dot{I}_j=\dot{R}_j=0$, finding that there is a disease free equilibrium (DFE) and an endemic equilibria (EE). Which of these equilibria is realized depends on the epidemiological value $R_0$, which we derive. We begin, however, with a discussion of the disease free equilibrium, which is characterized by the steady states $\bar{I}_j = \bar{I}_u = 0$. Solving under this condition gives us:
\begin{align}
    \bar{S}_j &= \frac{p(\bar{\Theta}_j)\bar{S}_u}{\lambda_j}, \label{DFE_Sj}\\
    \bar{S}_u &= \frac{\lambda_g\lambda_f}{\lambda_g\lambda_f+\lambda_fp(\bar{\Theta}_g)+\lambda_gp(\bar{\Theta}_f)}, \label{DFE_Su} \\
    \bar{R}_j &= \bar{R}_u = 0,\\
    \bar{\Theta}_g &= \frac{1}{\alpha c_g}\left((1-\alpha)(y_f-b) -\alpha c_f\bar{\Theta}_f - \frac{c_f(r+\lambda_f)}{q(\bar{\Theta}_f)}\right),\\
    \bar{\Theta}_f &= \frac{1}{\alpha c_f}\left((1-\alpha)(y_g-b+b_g) -\alpha c_g\bar{\Theta}_g - \frac{c_g(r+\lambda_g)}{q(\bar{\Theta}_g)}\right). 
\end{align}
from which we can derive the following equations:
\begin{align}
    \frac{\bar{F}\lambda_f}{p(\bar{\Theta}_f)} &= \frac{\bar{G}\lambda_g}{p(\bar{\Theta}_g)}, \label{eq_FG}\\
    \frac{c_f(r+\lambda_f)}{q(\bar{\Theta}_f)} - \frac{c_g(r+\lambda_g)}{q(\bar{\Theta}_g)} &= (1-\alpha)(y_f-y_g-b_g). \label{eq:Tightness_DFE}
\end{align}
If $b_g$ is large enough ($b_g \ge y_f-y_g),$ then \eqref{eq:Tightness_DFE} implies 
$\frac{c_f(r+\lambda_f)}{q(\Theta_f)} \le \frac{c_g(r+\lambda_g)}{q(\Theta_g)},$ i.e. the average hiring cost is higher in the gig labour market than in the formal labour market at the disease-free equilibrium. In the case where the separation rate is the same in both labour markets ($\lambda_g = \lambda_f),$ this implies $q(\Theta_f)<q(\Theta_g)$ and thus that labour market tightness is greater in the formal labour market. In turn, we have $\bar{F} > \bar{G}.$ If $b_g$ is small $(b_g < y_f-y_g)$, then the cost of hiring is higher in the formal labour market, $\frac{c_f(r+\lambda_f)}{q(\Theta_f)} \ge \frac{c_g(r+\lambda_g)}{q(\Theta_g)}.$ However, this does not allow us to make any determination on the relative sizes of $\bar{F}$ and $\bar{G}.$

Next consider the case where the system is at an endemic equilibrium (EE). Solving for the equilibria of the epidemiological equations of motion, we have the steady states:
\begin{align}
    \bar{S}_j &= \frac{(\gamma + \lambda_j)\bar{I}_j}{\beta_j \bar{I}}, \\
    \bar{S}_u &= \frac{ \gamma - \beta_f\bar{S}_f-\beta_g\bar{S}_g}{\beta_u},\\
    \bar{I}_j &= \frac{\lambda_j\bar{R}_j + \rho \bar{R}_j -p(\bar{\Theta}_j)\bar{R}_u}{\gamma}\\
    \bar{I}_u &=\frac{\rho}{\gamma}\bar{R} -\bar{I}_f-\bar{I}_g \\
    \bar{R}_j &= \frac{\beta_j\bar{S}_j\bar{I}-\lambda_j\bar{S}_j+p(\bar{\Theta}_j)\bar{S}_u}{\rho}, \\
    \bar{R}_u &= 1-\bar{S}-\bar{I}-\bar{R}_f-\bar{R}_g\\
    \bar{\Theta}_g &= \frac{1}{\alpha c_g}\left((1-\alpha)(y_f(1-\bar{i}_f) - b_u) -\alpha c_f\bar{\Theta}_f - \frac{c_f(r+\lambda_f)}{q(\bar{\Theta}_f)}\right),  \\
    \bar{\Theta}_f &= \frac{1}{\alpha c_f}\left((1-\alpha)(y_g(1-\bar{i}_g) + b_g - b_u)- \alpha c_g\bar{\Theta}_g - \frac{c_g(r+\lambda_g)}{q(\bar{\Theta}_g)}\right).
\end{align}
Equation \ref{eq_FG} still holds, and we have the equality
\begin{equation}
    \frac{c_f(r+\lambda_f)}{q(\bar{\Theta}_f)} - \frac{c_g(r+\lambda_g)}{q(\bar{\Theta}_g)} = (1-\alpha)(y_f(1-i_f)-y_g(1-i_g)-b_g).
\end{equation}

Note that we have assumed that unemployed infectious workers can still look for work. If they cannot, then unemployment is guaranteed to increase from the DFE equilibrium to the EE, driven by increased friction from infected unemployed unable to seek employment. However, if infected workers can search for work (as in our model), unemployment could theoretically decrease.

We can use the next-generation matrix method \citep{diekmann10} to determine the basic reproductive number. Identifying $I_f,I_g,I_u$ as disease variables, the flow into disease compartments is $\mathcal{F} = [\beta_fS_fI,\beta_gS_gI,\beta_uS_uI]^T$. The flow between disease compartments or away from them is $\mathcal{V} = [-(\lambda_f+\gamma) I_f,-(\lambda_g+\gamma)I_g,\lambda_f I_f + \lambda_g I_g - \gamma I_u]^T.$ Calculating the Jacobians $D\mathcal{F}$ and $D\mathcal{V}$ and substituting in the DFE from Equations \ref{DFE_Sj} and \ref{DFE_Su}, the next generation matrix is
\begin{equation}
    (D\mathcal{F})(D\mathcal{V})^{-1} = -\frac{\lambda_{f} \lambda_{g}}{\lambda_{f} \lambda_{g} + \lambda_{f} p{\left(\Theta_{g} \right)} + \lambda_{g} p{\left(\Theta_{f} \right)}}\begin{bmatrix}
    \frac{\beta_{f} p{\left(\Theta_{f} \right)}}{\gamma \lambda_{f}} & \frac{\beta_{f} p{\left(\Theta_{f} \right)}}{\gamma \lambda_{f}} & \frac{\beta_{f} p{\left(\Theta_{f} \right)}}{\gamma \lambda_{f}}\\\frac{\beta_{g} p{\left(\Theta_{g} \right)}}{\gamma \lambda_{g}} & \frac{\beta_{g} p{\left(\Theta_{g} \right)}}{\gamma \lambda_{g}} & \frac{\beta_{g} p{\left(\Theta_{g} \right)}}{\gamma \lambda_{g}}\\\frac{\beta_{u}}{\gamma} & \frac{\beta_{u}}{\gamma} & \frac{\beta_{u}}{\gamma}\end{bmatrix}
\end{equation}
where $\bar\Theta_j$ is the equilibrium value of the $j$the market tightness when $I=0$. The spectral radius of this matrix is the basic reproductive number, 
\begin{equation}
    R_0 =  \frac{\beta_{f} \lambda_{g} p{\left(\Theta_{f} \right)} + \beta_{g} \lambda_{f} p{\left(\Theta_{g} \right)} + \beta_{u} \lambda_{f} \lambda_{g}}{\gamma \left(\lambda_{g} p{\left(\Theta_{f} \right)}+\lambda_{f} p{\left(\Theta_{g} \right)} + \lambda_{f} \lambda_{g} \right)}.
\end{equation}
If $R_0>1$, then the solutions will converge to the endemic equilibrium. If $R_0<1$, then the solutions will converge to the disease-free equilibrium. Notably, if $\beta_u>\gamma,$ we have $R_0 > 1$
since we assume $\beta_f>\beta_g>\beta_u.$ I.e., if the transmission rate to the unemployed population is greater than the recovery rate, then the solutions are guaranteed to converge to the endemic equilibrium.

\subsection{Model calibration} \label{sec:calibration}

To understand short and medium-term dynamics of the system along with the effects of varying parameter values, we performed numerical simulations of the system. Table \ref{tbl:param} details the parameter values used for these simulations. Consider first the economic parameters. Workers' bargaining power and the matching function elasticity are set to $\alpha=\eta=0.5$, consistent with the Hosios condition for efficient labour market matching \citep{hosios90}. The job loss rate in the formal economy $\lambda_f$ is derived from assuming that jobs last on average $2.5$ years, which is in agreement with the literature \citep{hornstein05,shimer05}. Many gig jobs are held for only $3$ or fewer months, though some last as long as a year \citep{farrell18}. $\lambda_g$ is thus determined by assuming gig jobs are held for $6$ months on average. The matching function constant $\mu$ is taken from  taken from \cite{shimer05} and converted to a daily rate. The unemployment benefit $b_u$, which assumes that pre-epidemic daily wages in the formal market are normalized to one, is also taken from \cite{shimer05}. Assuming annual interest rates are approximately $5$\%, we set $r = \ln(1.05)/365 \approx 0.0001337$. Gig and other contingent employees can earn less than non-contingent workers \citep{banerjee23}, thus we assume that gig wages are $85$\% that of formal wages pre-epidemic (i.e.\ $W_g = 0.85$). Gig workers, however, are often willing to give up a sizable proportion of their wage for the flexibility gig work provides. This can be as high as $40$\% for Uber drivers \citep{chen19}. On the other hand, \cite{mas17} have shown that employees are willing to lose a more modest $8$\% of wages to work from home and schedule flexibility. Here we assume that $b_g=0.2$, $20$\% of pre-pandemic equilibrium wages of employees in the formal economy.

With these parameters and the free-entry conditions, $c_f$ and $\theta_g$ can be parameterized assuming the economy is in equilibrium with $\theta_f=1$:
\begin{align}
    c_f &= \frac{\mu(W_g + b_g - W_f)}{\lambda_g - \lambda_f}, \\
    \theta_g &= \sqrt[\eta]{\frac{c_f}{c_g}}.
\end{align}
Worker productivity in each economy and the vacancy cost of gig work can then be found using the wage equations  (Equations \ref{wage_eqn_formal} and \ref{wage_eqn_gig}).

The epidemiological parameters ($\beta_j$, $\gamma$, and $\rho$) are based on those observed during the COVID-19 pandemic \citep{bobrovitz23,byrne20,goldberg21,hassan23}. We assume that gig work results in a $5$\% reduction in the transmission rate of the disease, and unemployment in a further $5$\% reduction. Hybrid work days have been shown to have a reduced transmission rate compared to anchor days of $11-20$\% \cite{jung25}. However, we also explore varying reductions in transmission rates.

\subsection{Transitional dynamics}

\begin{figure}[ht!]
    \centering
    \includegraphics[width=\textwidth]{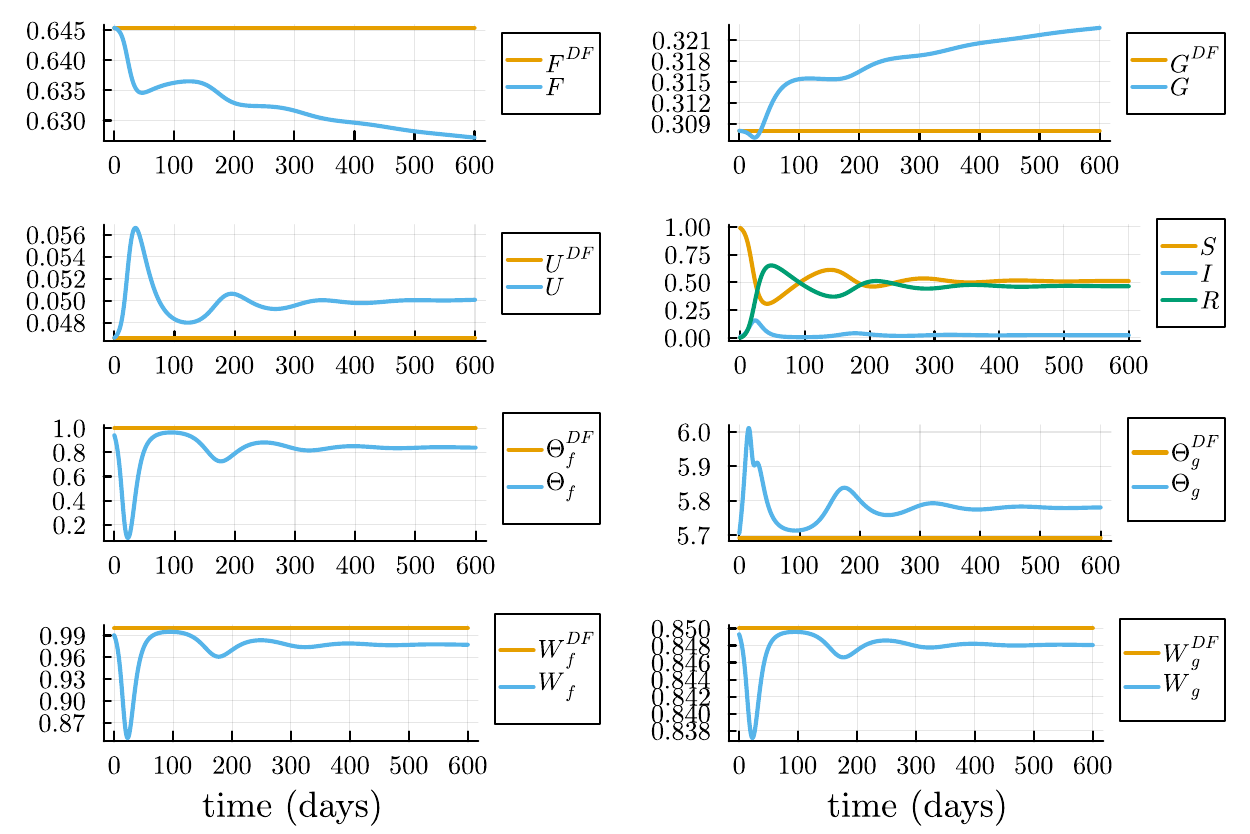}
    \caption{Time series of an epidemic in which $1\%$ of all workers are infectious. Plotting for comparison are the disease free (DF) equilibria for employment, unemployment, labour market tightnesses, and wages.}
    \label{fig:ts_EGUSIRThetaW}
\end{figure}

Figure \ref{fig:ts_EGUSIRThetaW} depicts the time series for a scenario where $1\%$ of the population across employment statuses is initially infected with the disease. Shown for comparison are the steady states for the disease free equilibrium. Employment in the formal and gig economies takes a prolonged period to equilibrate relative to the disease dynamics. Note that there is a secondary wave of infections that occurs several months after the initial outbreak. This secondary wave corresponds to fluctuations in the economic dynamics. As infections subside, workers return to the formal economy, which kindles a second wave of infections due to the higher transmission rate among workers employed in the formal economy. In response, there are also fluctuations in the number of susceptible and recovered individuals. Such swift rebounds in the employment rate, occurring over months, have been observed in previous studies of the COVID-19 epidemic \citep{cajner20,yu24}.

Returning to the economic dynamics, unemployment initially jumps as job separations outpace new job matches, but falls and equilibrates as the endemic equilibrium is reached. The initial shock of the epidemic leads to reductions in the numbers of workers in both economies. Then, the number of workers in the traditional and gig economies falls and rises, respectively. Since, gig workers are less often infected, which increases their net productivity. In the short run, job gains in the gig economy can offset job loses from the formal economy resulting in unemployment approaching pre-epidemic levels. However, in the long run, job losses from the formal economy outweigh the job gains in the gig economy, and thus unemployment increases relative to the disease free value.

We can see that initially there are large shocks in labour market tightness for both markets, which move in opposite directions. As the disease spreads, we observe a steep decline in labour market tightness of the formal labour market and a steep increase in the gig labour market. Long-term, these fluctuations diminish and reach the steady state where the formal and gig labour market tightness are below and above, respectively, their disease free values. However, the short-term transient shocks in unemployment and labour market tightnesses underscore the importance of early public policy interventions.

Finally, wages are depressed due to the epidemic. We observe similar trajectories for wages in both economies: wages steeply decline initially followed by a period of relatively moderate fluctuations. In the long run, they equilibrate to a modest reduction of their pre-epidemic values.

\subsection{Exogenous effects and interventions}

Here we explore the effects of different exogenous effects and interventions on the model. These represent the effects of public policies that aim to mitigate the deleterious effects of the disease, namely to reduce unemployment and infections, as well as unintended effects. Specifically, we consider the impact of the epidemic on job loss rates that disproportionately affects the formal economy, and government benefits to unemployed and gig workers. We then compare the outcomes of these three effects to one another.

\begin{figure}[ht!]
    \centering
    \includegraphics[width=\textwidth]{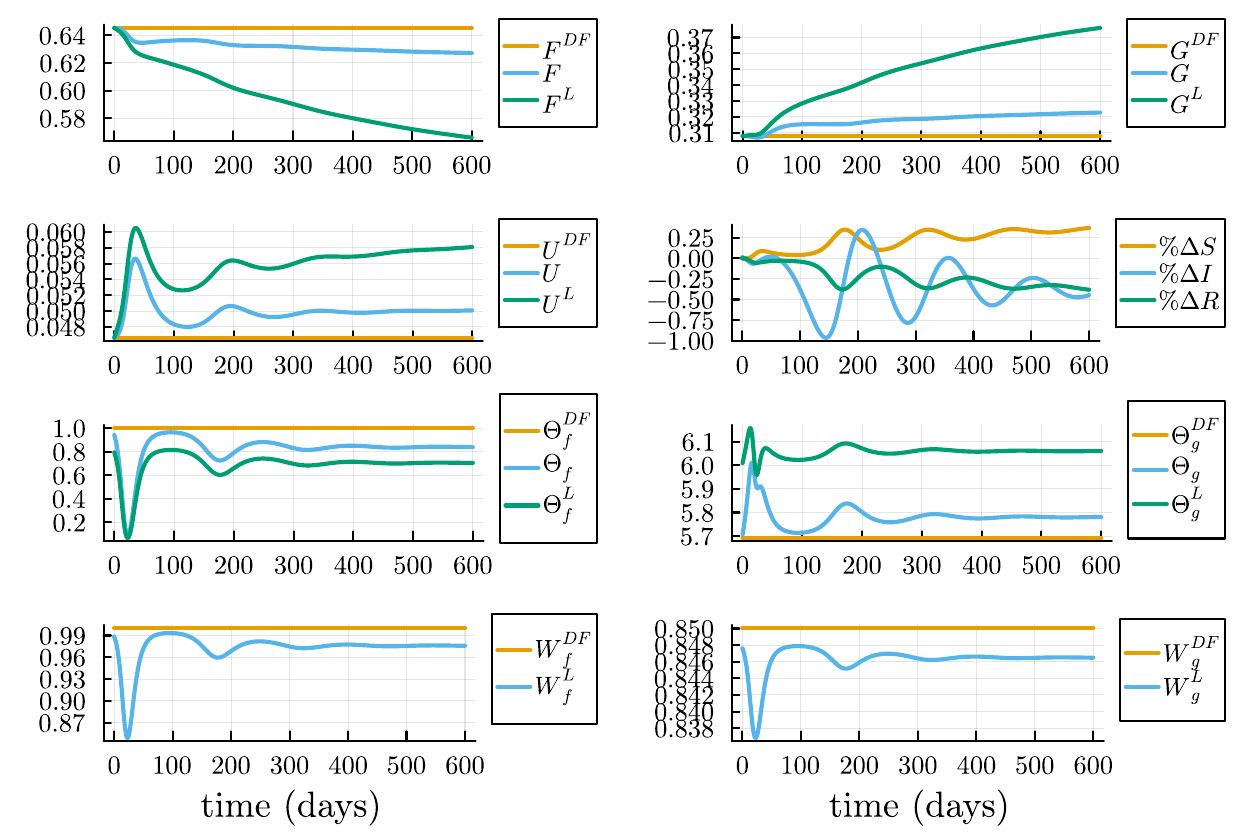}
    \caption{Time series of an epidemic where the job loss rate $\lambda_f$  for workers in the formal economy is $25$\% larger (results depicted with superscript $L$). Plotting for comparison are the time series for the endemic scenario without such an effect and the disease free scenario. Here $\% \Delta S$, $\% \Delta I$, and $\% \Delta R$ are the percentage changes between the epidemic scenarios with and without quarantine.}
    \label{fig:ts_quarantine}
\end{figure}

We begin by considering the impacts of an in increase in the job loss rate of employed workers while leaving the job loss rate of gig workers unaffected. This modelling choice represents the effects of prolonged quarantines and slow adaptations to the epidemic, which are are particularly relevant for formal employment. Job separation rates dramatically increased during the initial period of the COVID-19 pandemic. However, later rates in 2022 were approximately $25$\% those pre-pandemic and more recently have returned to approximately pre-pandemic levels \citep{huang24,BLS_JOLTS_Separations_2019_2024}. We thus explored the scenario of a $25$\% increase in $\lambda_f$, and depict the results in Figure \ref{fig:ts_quarantine}. We observe a decrease in employment in the formal economy while increasing both overall unemployment and employment in the gig economy. Infections are reduced, as we would expect given that the transmission rate is greater in the traditional economy. Thus, while this effect --- whether purposeful due to intermittent quarantines or not --- does reduce transmission of the disease and endemic infection levels, it increases unemployment and disproportionately impacts employment in the formal economy.

\begin{figure}[ht!]
    \centering
    \includegraphics[width=\textwidth]{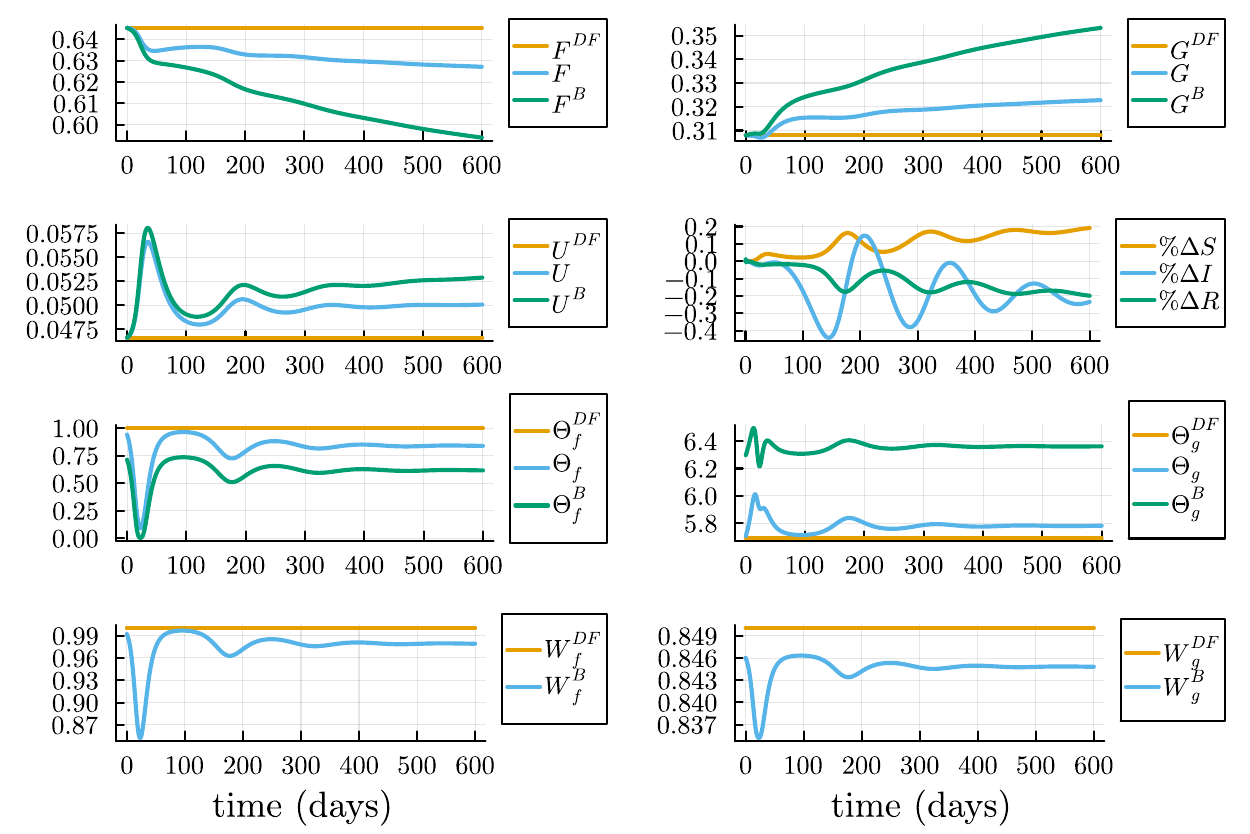}
    \caption{Time series of an epidemic where gig workers are provided a benefit $b_g=0.21$ to supplement their income (results depicted with superscript $B$). Plotting for comparison are the time series without such a benefit with disease and disease free scenarios. Here $\% \Delta S$, $\% \Delta I$, and $\% \Delta R$ are the percentage change between the epidemic scenarios where $b_g=0.21$ and $0.2$.}
    \label{fig:ts_gig_benefits}
\end{figure}

Next, we consider government provided benefits to gig workers derived from government programs that improve the viability of gig work such as Medicaid, which provides health insurance, and the Affordable Care act, which increased affordability and access to healthcare in the United States thus softening the ``job lock" effect of employer funded healthcare \citep{bailey17,even19}. We model the benefit of such programs to aide gig workers as an exogenous $5$\% increase in the benefit $b_g$. The impact of these changes is depicted in Figure \ref{fig:ts_gig_benefits}. We observe that this policy causes the employment in the formal economy to drop, yet also significant growth in employment in the gig economy. Infection levels are also reduced to a greater level than the scenario of increased separation rates for formal employment. 

\begin{figure}[ht!]
    \centering
    \includegraphics[width=\textwidth]{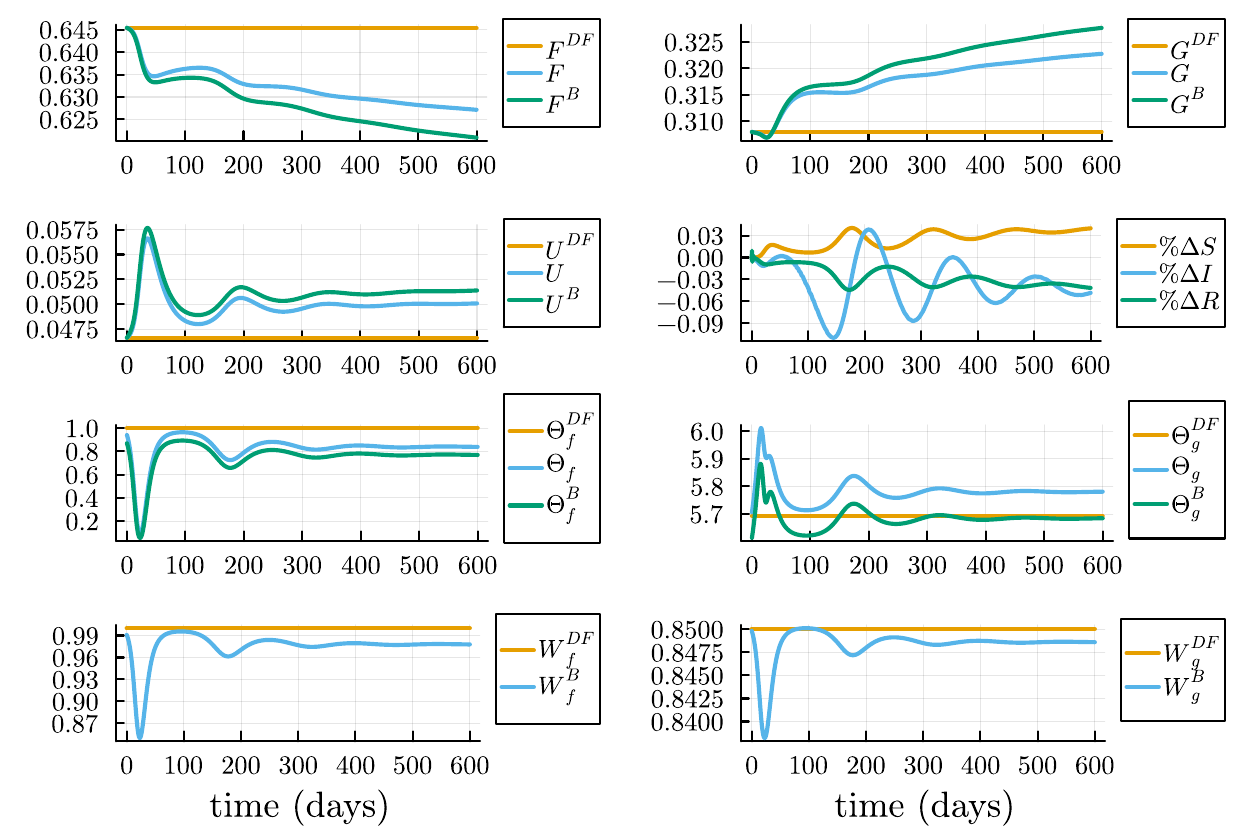}
    \caption{Time series of an epidemic where gig workers are provided a benefit $b_u=0.42$ to supplement their income ($F^B$ etc.). Plotting for comparison are the time series without such a benefit with disease and disease free scenarios. Here $\% \Delta S$, $\% \Delta I$, and $\% \Delta R$ are the percentage change between the epidemic scenarios where $b_g=0.21$ and $0.2$.}
    \label{fig:ts_bu}
\end{figure}

Finally, we consider the effects of a $5$\% increase in unemployment benefits in Figure \ref{fig:ts_bu}. This policy can be justified on the grounds that transmission rates are lower for the unemployed and thus overall infections may decrease. Our results confirm that this does decrease infections; however, comparing these results to the case of increased gig benefits, increased gig benefits resulted in a greater decrease in infections. Although, overall unemployment is higher due to gig benefit due to their effects on the formal economy.

\section{Discussion}

Our model contributes to the literature on labour market and epidemiological interactions, and dual labour markets of formal and informal, and high and low skilled work more generally \citep{gautier02, zylberstajn15, huang20}. Our approach differs from prior works such as \cite{kapicka22} and \cite{jackson24}, which focus on a single-sector labour market, by considering two labour markets, gig and formal. Additionally, we relax the assumption of lasting immunity post-recovery, which enables us to explore the long-term economical impacts of an endemic disease. If $R_0>1$, we show that an endemic equilibrium is reached relatively quickly compared to the economic system. However, this can result in a resurgence of infections as workers returning to the formal economy become infected and transmission rates increase.

We also consider the effects of exogenous factors on the outcomes of our model with an emphasis on mitigating economic and epidemiological trade-offs. We show a trade-off between unemployment and reducing infections for public policies of increasing unemployment benefits or authorizing quarantines. Echoing research showing that changes to the gig economy were strongly driven by policies rather than the disease itself \citep{cao22}, we find that unemployment is increased more by these policies than the disease itself. However, we observed minimal impact on unemployment when benefits are provided to gig workers while also mitigating the spread of disease. Our results thus tie into broader research on optimal quarantine and testing policies \citep{alvarez21, berger22, piguillem22}, and the effect of unemployment insurance \citep{marinescu21} and short-term work schemes to preserve employment \citep{osuna22} during an epidemic. 

There are several limitations of this study, such as a homogeneous labour force and no regional variations in the economy. Homogeneity in the gig economy is particularly relevant. We assumed that transmission rates were lower in the gig economy than the formal economy due to work being independent and relatively infrequent. However, transport gig work may pose a high risk of infection \citep{lew21,anand22}, and thus the type of gig work may be highly important to model. In the long-term, technological innovations, which have greatly affected the gig economy \citep{liu24}, could also have significant impacts on these results. E.g.,\ work from home options for formal employment could reduce the chances of infection. However, this can result in productivity loss \citep{atkin23,gibbs23,emanuel24}. Future research could explore these complexities within the framework presented here. We also have not considered broader impacts of increases in gig work. On the one hand, the gig economy has been shown to reduce customer satisfaction \citep{shin23}, and on the other hand it has had an empowering effect on workers \citep{liz21}. Thus the effects of spreading disease and these policies to mitigate it are deeper than those considered here.

\section*{Declarations}
\subsection*{Code and Data Availability}
Code to run the simulations is available at github.com/bmorsky/epi-econ.
\subsection*{Competing interests}
The authors have no relevant financial or non-financial interests to disclose.
\subsection*{Funding}
No funding was received for conducting this study.
\bibliography{epi_econ}
\bibliographystyle{apalike}

\end{document}